\documentclass{article}
\usepackage{graphicx}
\usepackage{amsmath}

\input{tcilatex}

\begin{document}

\title{Magnetoresistance in the s-d Model with Arbitrary Impurity Spin}
\author{Ding Kaihe$^{\ast }$ and Bao-Heng Zhao$^{\ast \ast }$ \\
Department of Physics, Graduate School\\
Chinese Academy of Sciences\\
P.O. Box 3908, Beijing 100039, China\\
Email: $^{\ast }$ Dingkaih@mails.gscas.ac.cn; \\
$^{\ast \ast }$ Zhaobh@gscas.ac.cn}
\date{}
\maketitle

\begin{abstract}
The magnetoresistance, the number of the localized electrons, and the s-wave
scattering phase shift at the Fermi level for the s-d model with arbitrary
impurity spin are obtained in the ground state. To obtain above results some
known exact results of the Bethe ansatz method are used. As the impurity
spin S = 1/2, our results coincide with those obtained by Ishii \textit{et al%
}. The compairsion between the theoretical and experimental
magneticresistence for impurity S = 1/2 is re-examined.

PACS numbers: 75.20Hr, 72.15Qm, 75.30Hx.
\end{abstract}

{\large 1. Introduction}

\bigskip The magnetoresistance for the impurity spin S = 1/2 at temperature $%
T=0$ was obtained in the s-d model by many authors [1]-[4]. For arbitrary
impurity spin $j$ at $T=0$ the magnetoresistance was obtained in the
Coqblin-Schrieffer model [5] (for reviews see [6] and [7]). The purpose of
this paper is to study the \ magnetoresistance for\ higher impurity spin in
the s-d model in the ground state.\ The electric resistivity obtained by us
are given below in (11)-(12). For impurity spin $S=$ $1/2$, (12) coincides
with the known result [1]-[4]. We also make a comparison of the electric
resistivity for $S=1/2$ with experimental data obtained by Felsh \textit{et
al.} [8]. The same comparison was also done by [5]. Our result is different
from that of [5] as shown below in Fig. 1.

{\large 2. Numbers of localized electrons, scattering phase shifts and
magnetoresistance}

The s-d exchange Hamiltonian for an impurity spin $S$ localized at the
origin is 
\begin{equation}
\hat{H}=\sum_{\mathbf{k,}\sigma }\varepsilon _{k}c_{\mathbf{k}\sigma
}^{\dagger }c_{\mathbf{k}\sigma }+\frac{J}{2}\sum_{\mathbf{k,k}^{\prime
},\sigma ,\sigma ^{\prime }}c_{\mathbf{k}\sigma }^{\dagger }\mathbf{\sigma }%
_{\sigma \sigma ^{\prime }}c_{\mathbf{k}^{\prime }\sigma ^{\prime }}\cdot 
\mathbf{S,}  \tag{1}
\end{equation}
where $c_{\mathbf{k}\sigma }$ ($c_{\mathbf{k}\sigma }^{\dagger }$) is the
annihilation (creation) operator for an electron with wave vector $\mathbf{k}
$ and spin $\sigma $, $\varepsilon _{k}$ is the kinetic energy of the
electron with wave vector $\mathbf{k}$, $\mathbf{S}$ is the impurity spin, $%
\mathbf{\sigma }$ is the Pauli matrix, and $J$ is the coupling constant. We
will only consider the case of antiferromagnetic coupling ($J>0$). We can
also introduce a magnetic field $H$, then a Zeeman term should be added to
the right hand side of (1).

In the ground state, as the impurity spin $S=1/2$, Yosida and Yoshimori
calculated the numbers of the localized electrons around the impurity.
Through the Friedel sum rule they obtained the scattering phase shift at the
Fermi level and then the magnetoresistance [3] (see also a more succinct \
presentation [9]). In this work we will generalize the study of Yosida and
Yoshimori to the case of arbitrary impurity spin. We assume that certain
results of the Bethe ansatz method in one dimensional space [10] can be
applied \ to the s-wave electrons with the above Hamiltonian (1). Under this
assumption, the deduction is dramatically simplified.

If the magnetic field $H=0,$ Mattis pointed out [11], and Fateev and
Wiegmann proved [12], using the Bethe ansatz method, that the spin of the
system of the s-d model equals $S-1/2$ in the ground state. The electrons
interacting with the impurity in the s-wave can be considered as an one
dimensional system. The total spin of the electrons, in the ground state,
according to a theorem of Lieb and Mattis [13], takes the lowest possible
value. If the number of electrons is odd, the total spin of the electrons
(not including impurity spin) is equal to $1/2$, which coupled with the
impurity spin leads to the total spin $S-1/2$ . The wave function of the
system can be written as 
\begin{equation}
\Psi =\psi _{-1/2}\chi _{S}+\;\psi _{1/2}\chi _{S-1},  \tag{2}
\end{equation}
where $\chi _{M}$ is the spin wave function of the impurity, and $M$ is the $%
z$ component of the spin of the impurity; $\psi _{m}$ is the wave function
of electrons, and $m$ is the $z$ component of the total spin of electrons.
The electrons away from the impurity are unpolarized, however, the localized
electrons are polarized. So that $m$\ is due to localized electrons. The
numbers of the localized electrons in $\Psi $ satisfy conditions [9]: 
\begin{eqnarray}
n_{\uparrow }^{(-1/2)} &=&n_{\uparrow }^{(1/2)}-1,  \TCItag{3} \\
n_{\downarrow }^{(-1/2)} &=&n_{\downarrow }^{(1/2)}+1\;  \notag
\end{eqnarray}
where $n_{\uparrow }^{(m)}$ ($n_{\downarrow }^{(m)}$) is the number of the
localized electrons with spin up (down) in $\psi _{m}.$ Condition (3) is the
consequence of Anderson's orthogonality theorem [14] and [9]. These
conditions can be understood as follows. In the spin flip scattering, $\psi
_{-1/2}\chi _{S}$ is transformed to $\psi _{1/2}\chi _{S-1},$ or vice versa,
in other words, $\psi _{-1/2}\chi _{S}$ and $\psi _{1/2}\chi _{S-1}$ are
connected by the s-d exchange interaction. The matrix element of the
interaction Hamiltonian between $\psi _{-1/2}\chi _{S}$ and $\psi _{1/2}\chi
_{S-1}$ nonvanished yields condition (3).\ 

Now, let a uniform magnetic field $H$ in $-\mathbf{\hat{z}}$ direction be
switched on. Of course, the electron wave functions $\psi _{\pm 1/2}$\ will
vary and deviate from their original forms, and the z-components of the
spins of the electron system will deviate from $\pm \frac{1}{2}$, however,
for convenience, we still denote them as $\psi _{\pm 1/2}$. Thus the
subscripts $\pm \frac{1}{2}$of $\psi _{\pm 1/2}$ do not indicate the
z-components of the spin of the electron system to be $\pm \frac{1}{2}$ if $%
H\neq 0$. Furthermore, we assume that there is no components with $\chi
_{M<S-1}$ being generated as $H\neq 0$, because smaller $M$ is corresponding
to higher Zeeman energy. So that we still have (2) and (3). Besides, we have
charge neutrality condition: 
\begin{equation}
(n_{\uparrow }^{(-1/2)}+n_{\downarrow }^{(-1/2)})P_{S}+(n_{\uparrow
}^{(1/2)}+n_{\downarrow }^{(1/2)})P_{S-1}=0,  \tag{4}
\end{equation}
where $P_{S}=\left| \left\langle \psi _{-1/2}\left| \psi _{-1/2}\right.
\right\rangle \right| ^{2}$ and $P_{S-1}=\left| \left\langle \psi
_{1/2}\left| \psi _{1/2}\right. \right\rangle \right| ^{2}$ are the
probability of the system in states $\psi _{-1/2}\chi _{S}$ \ and $\psi
_{1/2}\chi _{S-1},$ respectively. The wave functions $\Psi $ and $\chi _{M}$
have been normalized to unity, thus $P_{S}+P_{S-1}=1$. The charge neutrality
condition (4) can be justified by an inspection of the Bethe wave function
[15]and [16]. The Bethe wave function shows that the number of the
conduction electrons equals the total number of electrons, which implies
that the total charge of the localized electrons vanishes.

It is evident that 
\begin{equation}
SP_{S}+(S-1)P_{S-1}=\left\langle S_{i}^{z}\right\rangle  \tag{5}
\end{equation}
and 
\begin{equation}
\frac{1}{2}(n_{\uparrow }^{(-1/2)}-n_{\downarrow }^{(-1/2)})P_{S}+\frac{1}{2}%
(n_{\uparrow }^{(1/2)}-n_{\downarrow }^{(1/2)})P_{S-1}=\left\langle
S_{e}^{z}\right\rangle ,  \tag{6}
\end{equation}
where $\left\langle S_{i}^{z}\right\rangle $ ($\left\langle
S_{e}^{z}\right\rangle $) is the expectation value of the z-component of the
spin for the impurity (localized electrons) in the ground state.

Eqs.(3) and (4) lead to 
\begin{equation}
n_{\uparrow }^{(-1/2)}=-n_{\downarrow }^{(-1/2)},\;\;n_{\uparrow
}^{(1/2)}=-n_{\downarrow }^{(1/2)}.  \tag{7}
\end{equation}
Let us denote the z-component of localized spin as $M_{i}\equiv \left\langle
S_{i}^{z}\right\rangle +\left\langle S_{e}^{z}\right\rangle .$ Eqs. (5 ) -
(7) and (3) lead to 
\begin{equation}
M_{i}=n_{\uparrow }^{(-1/2)}+S,  \tag{8}
\end{equation}
and then 
\begin{eqnarray}
n_{\uparrow }^{(-1/2)} &=&-n_{\downarrow }^{(-1/2)}=M_{i}-S,  \TCItag{9} \\
n_{\uparrow }^{(1/2)} &=&-n_{\downarrow }^{(1/2)}=M_{i}-S+1.  \notag
\end{eqnarray}

From the Friedel sum rule [9,17] we immediately obtain the scattering phase
shifts of the s-wave at the Fermi level: 
\begin{eqnarray}
\delta _{\uparrow }^{(-1/2)}(\varepsilon _{F}) &=&-\delta _{\downarrow
}^{(-1/2)}(\varepsilon _{F})=\pi (M_{i}-S),  \TCItag{10} \\
\delta _{\uparrow }^{(1/2)}(\varepsilon _{F}) &=&-\delta _{\downarrow
}^{(1/2)}(\varepsilon _{F})=\pi (M_{i}-S+1)  \notag
\end{eqnarray}
where $\delta _{\sigma }^{(m)}(\varepsilon _{F})$ is the scattering phase
shift of the electron in state $\psi _{m}$ with z-component of spin $\sigma $%
.

From (10) we obtain the impurity magnetoresistance:

\begin{equation}
R(H)=R_{0}\sin ^{2}(\pi M_{i}),\;\;S=\text{ integer,}  \tag{11}
\end{equation}
and 
\begin{equation}
R(H)=R_{0}\cos ^{2}(\pi M_{i}),\;\;S=\text{ half-integer,}  \tag{12}
\end{equation}
where $R_{0}=R(H=0)$ (remember $M_{i}(H=0)=S-1/2$ ). When $S=1/2$, (12)
coincides with those obtained in [1]-[3] with $M_{i}$ replaced by an
approximate expression. The same expression of (12) for $S=1/2$ was also
given by [4].

{\large 3. Comparison with experimental data for }$S=1/2$

We compare the theoretical result with experimental data in Fig.1. The solid
curve for magnetoresistance $R(H)$ versus $H$ for $S=\frac{1}{2}$ is
obtained from (12). The experimental data \ for the magnetoresistance of
(La,Ce)Al$_{2}$ are taken from [8]. The dashed curve is obtained from (12),
but takes the magnetization $\mathcal{M}^{i}$\ from (9.25) of [15] of Andrei 
\textit{et al.,} and it is the same curve shown by Schlottmann in [5]. Fig.2
gives the magnetoresistance as a function of the magnetic field for
impurities $S=1/2,1,3/2$ obtained from (11) and (12). To compute $R(H)$
(except those for the dashed curve in Fig.1) we have used\ the exact
expression of $M_{i}$ obtained by the Bethe ansatz method as follows, 
\begin{eqnarray}
M_{i}(g\mu _{B}H &>&2k_{B}T_{H})=S-\frac{1}{2\pi ^{3/2}}\int_{0}^{\infty
}d\omega \frac{\sin (2\pi \omega S)}{\omega }\Gamma (1/2+\omega )(\frac{%
\omega }{e})^{-\omega }  \notag \\
&&\times \exp [-2\omega \ln (\frac{g\mu _{B}H}{2k_{B}T_{H}})]  \TCItag{13}
\end{eqnarray}
\begin{eqnarray}
M_{i}(g\mu _{B}H &<&2k_{B}T_{H})=S-1/2  \TCItag{14} \\
&&+\frac{1}{2\pi ^{3/2}}\mathcal{P}\int_{0}^{\infty }d\omega \frac{\exp
(2\omega \ln \frac{g\mu _{B}H}{2k_{B}T_{H}})}{\omega }\Gamma (1/2-\omega
)\left( \frac{\omega }{e}\right) ^{\omega }\sin [2\pi (S-1/2)\omega ]  \notag
\\
&&+\sum_{n=0}^{\infty }\frac{(-1)^{n}(\frac{g\mu _{B}H}{2k_{B}T_{H}})^{2n+1}%
}{2\sqrt{\pi }n!(n+1/2)}\left( \frac{n+1/2}{e}\right) ^{n+1/2}\cos [2\pi
(S-1/2)(n+1/2)]  \notag
\end{eqnarray}
where $T_{H}=\frac{2N}{L}\sqrt{\frac{2\pi }{e}}e^{-\pi /g^{\prime }}$ with $%
g^{\prime }=\frac{1}{S+1/2}\tan [(S+1/2)J/2].$ (This $T_{H}\equiv (T_{H})_{%
\text{Wiegmann}}$ is defined as that in [16], not as $T_{H}\equiv (T_{H})_{%
\text{Andrei}}$ in [15]. The relation between them is $(T_{H})_{\text{%
Wiegmann}}=\sqrt{8}(T_{H})_{\text{Andrei}}$ for $J<<1.$)$\ N$ is the number
of electrons, $L$ is the length of the system, $\mu _{B}$ is the Bohr
magneton, and $g$ is the Land\'{e} g-factor.

We have to make some remarks.

1, Eqs. (13) and (14) are essentially the same as (31) and (33) of [12]
(where $g=2$, $\mu _{B}=1,\;k_{B}=1$), respectively, but with corrections:
(a) in (33) of [12] $\int_{0}^{\infty }d\omega ...$ is replaced by the
principal-value integral$\ \mathcal{P}\int_{0}^{\infty }d\omega ...$, and $%
\sum_{n=0}^{\infty }...$ is replaced by $\frac{1}{2}\sum_{n=0}^{\infty }...$%
, (b) in (33) of [12] the integrand multiplies a factor $\sin [2\pi
(S-1/2)\omega ],$ (c) in (31) and (33) of [12] $T_{K}$ is replaced by $T_{H}$%
. Corrections (b) and (c) have already been done by [16].

2, To derive (13) and (14), the Pauli susceptibility is assumed to be $\chi
=L/2\pi $ (for $\mu _{B}=1,\;g=2$) in one dimensional space which
corresponds to take the coupling constant $J=0^{+}$. This value of $\chi $
coincides with that adopted by Tsvelick and Wiegmann [16], but different
from that adopted by Andrei \textit{et al}. [15]. Andrei \textit{et al}.
assumed the value of $\chi $ corresponding to $J=0$. Since the density of
states at the Fermi level for $J=0$ is two times for $J=0^{+}$ in the Bethe
ansatz method, the Pauli susceptibility adopted by Andrei \textit{et al.}
equals $\chi =L/\pi .$ In our opinion, it is appropriate to take $J=0^{+}.$
If one assumes $J=0,$ as Andrei \textit{et al.}, the factor $\frac{g\mu _{B}H%
}{2k_{B}T_{H}}$ in (13) and (14) should be replaced by $\frac{g\mu _{B}H}{%
k_{B}T_{H}}.$

3, A similar comparison between the experimental magnetoresistance and the
theoretical result of Andrei for $S=$ $1/2$ [4] has been done by Schlottmann
[5]. \ However, what we do here is different form [5]. In [5], (a) the
experimental electric resistivity is less than those obtained by Felsh et
al.[8] about $0.6/\mu \;\Omega \;cm$ although [5] referred to [8], (b) to
compute $M_{i}$ Schlottmann[5] used the result of [15], where the Pauli
susceptibility $\chi $ corresponding to $J=0$. With corrections on these two
points, we see that the theoretical curve for $S=1/2$ fits the experimental
data well except at $H$ $=50$\ Oe.

{\large 4. Conclusion}

The magnetoresistance of s-d model with arbitrary impurity spin in the
ground state is obtained by combining the Yosida-Yoshimori method and
results from the Bethe ansatz. The comparison of the magnetoresistance of
the s-d model for $S=1/2$ in the ground state with experimental data is
re-examined. We have shown that one should use (13)-(14) to calculate the
magnetoresistance to fit the experimental data. As shown in Fig. 1, the
solid curve is closer to the experimental data than the older result (the
dashed curve).

{\large Acknowledgment}

We thank Professor G. Su for discussions on the Kondo problem. This work is
supported partially by NNSF of China.

{\large References}

[1] H. Ishii, Prog. Theor. Phys. \textbf{43}, 578 (1970).

[2] D. R. Hamann, Phys. Rev. \textbf{B2}, 1373 (1970).

[3] K. Yosida and A. Yoshimori, in \textit{Magnetism}, Vol. 5, edited by G.
Rado and H. Suhl (New York, Academic Press 1973).

[4] N. Andrei, Phys. Lett. \textbf{87A}, 299 (1982). See also [14].

[5] P. Schlottmann, Phys. Rev. \textbf{B35}, 5279 (1987).

[6] P. Schlottmann, Phys. Rep. \textbf{181}, 4 (1989).

[7] A. C. Hewson, \textit{The Kondo Problem to Heavy Fermions} (Cambridge,
Cambridge Univ. Press 1996).

[8] W. Felsch and K. Winzer, Solid State Commun. \textbf{13}, 569 (1973).

[9] K. Yosida, \textit{Theory of Magnetism} (Springer, Berlin 1996).

[10] V. A. Fateev and P. B. Wiegmann, Phys. Lett. \textbf{81 A}, 179 (1981);
Phys. Rev. Lett. \textbf{46}, 1595 (1981). See also A. M. Tsvelick and P. B.
Wiegmann, Adv. Phys. \textbf{32}, 453 (1983), and H. Furuya and J. H.
Lowenstein, Phys. Rev. \textbf{B 25}, 5935 (1982).

[11] D. Mattis, Phys. Rev. Lett. \textbf{19}, 1478 (1967).

[12] V. A. Fateev and P.B. Wiegmann, Phys. Lett. \textbf{81A}, 179 (1981).

[13] E. Lieb and D. Matts, Phys. Rev. \textbf{125}, 164 (1962).

[14] P. W. Anderson, Phys. Rev. Lett. \textbf{18}, 1049 (1967); Phys. Rev. 
\textbf{164}, 352 (1967).

[15] N. Andrei, H. Furuya, and J. H. Lowenstein, Rev. Mod. Phys. \textbf{55}%
, 331 (1983).

[16] A. M. Tsvelick and P. B. Wiegmann, Adv. Phys. \textbf{32}, 453 (1983).

[17] J. S. Langer and V. A. Ambegaokar, Phys. Rev. 121, 1090 (1961).

[18] V. T. Rajan, J. H. Lowenstein and N. Andrei, Phys. Rev. Lett. \textbf{49%
}, 497 (1982).

\bigskip

\bigskip {\large Figure Captions}

Fig.1 Magnetoresistance of (La, Ce)Al$_{2}$. The experimental data are taken
from [8]. The magnetoresistance $R(H)$ corresponding to $S=1/2$ shown by the
solid curve is obtained by using (12) - (14) with $T_{H}=\frac{1}{W}\sqrt{%
\frac{8\pi }{e}}T_{K},$ and the Wilson number $W=1.290265$ [15]. The Kondo
temperature $T_{K}=0.20K$ and $g=10/7$ are the same ones used by Rajan 
\textit{et al}. [18].\textit{\ }The dashed curve is obtained by using $%
\mathcal{M}^{i}$\ of (9.25) in [15] of Andrei \textit{et al., }who assumed
the Pauli susceptibility corresponding to $J=0$, and it is just the curve of
Schlottmann [5]. \textit{\ \ \ \ \ \ \ \ \ \ \ \ \ \ \ \ \ \ \ \ \ \ \ \ \ \
\ \ \ \ \ \ \ \ \ \ \ \ \ \ \ \ \ \ \ \ \ \ \ \ \ \ \ \ \ \ \ \ \ \ \ \ \ \
\ \ \ \ \ \ \ \ \ \ \ \ \ \ \ \ \ \ \ \ \ \ \ \ \ \ \ \ \ \ \ \ \ \ \ \ \ \
\ \ \ \ \ \ \ \ \ \ \ \ \ \ \ \ \ \ \ \ \ \ \ \ \ \ \ \ \ \ \ \ \ \ \ \ \ \
\ \ \ \ \ \ \ \ \ \ \ \ \ \ \ \ \ \ \ \ \ \ \ \ \ \ \ \ \ \ \ \ \ \ \ \ \ \
\ \ \ \ \ \ \ \ \ \ \ \ \ \ \ \ \ \ \ \ \ \ \ \ \ \ \ \ \ \ \ \ \ \ \ \ \ \
\ \ \ \ \ \ \ \ \ \ \ \ \ \ \ \ \ \ \ \ \ \ \ \ \ \ \ \ \ \ \ \ \ \ \ \ \ \
\ \ \ \ \ \ \ \ \ \ \ \ \ \ \ \ \ \ \ \ \ \ \ \ \ \ \ \ \ \ \ \ \ \ \ \ \ \
\ \ \ \ \ \ \ \ \ \ \ \ \ \ \ \ \ \ \ \ \ \ \ \ \ \ \ \ \ \ \ \ \ \ \ \ \ \
\ \ \ \ \ \ \ \ \ \ \ \ \ \ \ \ \ \ \ \ \ \ \ \ \ \ \ \ \ \ \ \ \ \ \ \ \ \
\ \ \ \ \ \ \ \ \ \ \ \ \ \ \ \ \ \ \ \ \ \ \ \ \ \ \ \ \ \ \ \ \ \ \ \ \ \
\ \ \ \ \ \ \ \ \ \ \ \ \ \ \ \ \ \ \ \ \ \ \ \ \ \ \ \ \ \ \ \ \ \ \ \ \ \
\ \ \ \ \ \ \ \ \ \ \ \ \ \ \ \ \ \ \ \ \ \ \ \ \ \ \ \ \ \ \ \ \ \ \ \ \ \
\ \ \ \ \ \ \ \ \ \ \ \ \ \ \ \ \ \ \ \ \ \ \ \ \ \ \ \ \ \ \ \ \ \ \ \ \ \
\ \ \ \ \ \ \ \ \ \ \ \ \ \ \ \ \ \ \ \ \ \ \ \ \ \ \ \ \ \ \ \ \ \ \ \ \ \
\ \ \ \ \ \ \ \ \ \ \ \ \ \ \ \ \ \ \ \ \ \ \ \ \ \ \ \ \ \ \ \ \ \ \ \ \ \
\ \ \ \ \ \ \ \ \ \ \ \ \ \ \ \ \ \ \ \ \ \ \ \ \ \ \ \ \ \ \ \ \ \ \ \ \ \
\ \ \ \ \ \ \ \ \ \ \ \ \ \ \ \ \ \ \ \ \ \ \ \ \ \ \ \ \ \ \ \ \ \ \ \ \ \
\ \ \ \ \ \ \ \ \ \ \ \ \ \ \ \ \ \ \ \ \ \ \ \ \ \ \ \ \ \ \ \ \ \ \ \ \ \
\ \ \ \ \ \ \ \ \ \ \ \ \ \ \ \ \ \ \ \ \ \ \ \ \ \ \ \ \ \ \ \ \ \ \ \ \ \ }

\bigskip

Fig.2 The magnetoresistance as a function of the magnetic field obtained
from (11)-(14) for impurities $S=1/2,1,3/2$. The parameters are the same as
those in Fig.1.

\end{document}